# Towards Run Time Estimation of the Gaussian Chemistry Code for SEAGrid Science Gateway

Angel Beltre
Department of Computer Science
Binghamton University SUNY
Binghamton, New York, U.S.A.
abeltre1@binghamton.edu

Shehtab Zaman
Department of Computer Science
Binghamton University SUNY
Binghamton, New York, U.S.A.
szaman5@binghamton.edu

Kenneth Chiu
Department of Computer Science
Binghamton University SUNY
Binghamton, New York, U.S.A.
kchiu@binghamton.edu

Sudhakar Pamidighantam
Science Gateways Research Center, Pervasive technology Institute and Department of Chemistry
Indiana University
Bloomington, Indiana, U.S.A.
pamidigs@iu.edu

Xingye Qiao
Department of Mathematical Sciences
Binghamton University SUNY
Binghamton, New York, U.S.A.
qiao@math.binghamton.edu

Madhusudhan Govindaraju
Department of Computer Science
Binghamton University SUNY
Binghamton, New York, U.S.A.
mgovinda@binghamton.edu

## ABSTRACT

Accurate estimation of the run time of computational codes has a number of significant advantages for scientific computing. It is required information for optimal resource allocation, improving turn-around times and utilization of science gateways. Furthermore, it allows users to better plan and schedule their research, streamlining workflows and improving the overall productivity of cyberinfrastructure. Predicting run time is challenging, however. The inputs to scientific codes can be complex and high dimensional. Their relationship to the run time may be highly non-linear, and, in the most general case is completely arbitrary and thus unpredictable (i.e., simply a random mapping from inputs to run time). Most codes are not so arbitrary, however, and there has been significant prior research on predicting the run time of applications and workloads. Such predictions are generally application-specific, however. In this paper, we focus on the Gaussian computational chemistry code. We characterize a data set of runs from the SEAGrid science gateway with a number of different studies. We also explore a number of different potential regression methods and present promising future directions.

## 1 INTRODUCTION

One of the main challenges in HPC environments, accessed via science gateways, is determining the amount of computational resources needed for the cost-efficient execution of scientific workflows. This particular challenge takes the form of minimizing the makespan of the current job set. Such minimization requires the ability to accurately estimate how long jobs will run. In addition to overall system optimization, accurate run time estimation benefits individual users in a number of ways. For example, a user may want to allocate more resources for a job that will run longer, perhaps at a higher cost. Likewise, in some instances, a longer execution time may be tolerable, in which case the user may wish to reduce resources, which may also reduce the cost. In order to decide, however, the user needs an accurate estimate of the run time. Execution time estimation is a well studied problem for various scientific and large scale software similar to Gaussian [8] [11] [12] [20]. Execution time research has been done using hardware performance counters, statistical and algorithmic analysis, and more recently various machine learning approaches. For example, some research seeks to investigate how a particular code will execute on different hardware. Increased use of shared HPC and cloud [23] [24] computational services has led to higher availability of data, suggesting machine learning approaches as a viable option for execution time prediction.

In this paper, we explore learning models based on Gaussian application input parameters to derive the run time prediction. We assume that the architectural variations will also affect run time, but are left for the models to recognize this natively from the dataset. We used a data set of runs available from the SEAGrid science gateway [7]. There are about 783 registered users in the gateway and for the period 2016-2018, the gateway served 94,868,631 XSEDE services units (26,091,516 CPU hrs) running 77,464 jobs. During this period the community published 52 peer reviewed publications citing the gateway for resources and services. Our study of the input route keywords, system, and derived features shows that a full assessment of such input features is critical to enable the best way to answer fundamental questions when predicting run times. This paper is organized as follows. In section 2 we provide a high level background of the science and the software and its input features. The data used in the learning and its processing is described in section 3. In section 4 a statistical analysis of data is provided to explore the component representations. Regression based learning is presented in section 5. Section 6 provides related work and a discussion of current and remaining challenges is presented in 7. The the paper is concluded in section 8 with a conclusion and outlook.

## 2 BACKGROUND
## 2.1 High level background

Quantum chemistry (QC) techniques provide a way to compute the energetics of molecular systems at unit kcal/mole accuracy, which is critical in applications such as drug design, material design and discriminating analysis of natural phenomena at the molecular level. Such computations are also embedded as part of a workflow that provides parameters for lower level empirical force field (FF) models or integrated into complex layered models that combine QC models with parameterized QC or empirical FF models to either speedup the runs or address larger systems. The requirement to estimate run time for such scenarios is absolutely essential to ensure that the compute allocations are efficiently used by the individual researcher and by the community in a community oriented infrastructure such as a science gateway. For the workflow a sum total run time (over all tasks) needs to be requested and an estimate for individual tasks is needed for both validation of the resource availability as well as efficient processing. The usual gross over-estimation is detrimental in these scenarios as the execution may fail to be launched due to larger than needed request or delayed due to the time needed for reserving resources that are at the end may not be needed/used.

## 2.2 Description of Gaussian Software [9]

Here we provide only a very brief introductory description of the algorithmic operation of the Gaussian software just to provide a glimpse of the complexity as an elaborate one is out of scope. Gaussian software provides two ways to describe molecular systems and compute their properties. 1. It solves Schrodinger's equation (SE) for molecular systems for an optimized electronic wavefunction and the total energy of the system. 2. Alternatively, it optimizes electron density by solving the Kohn-Sham equations. The density can be constructed from wavefunctions to start the initial guess density. Various methods provide approximations to this calculation as the analytical solutions for SE do not exist beyond the simplest of systems such as dihydrogen molecule. An iterative approach is adopted to self consistently update a guess wavefunction (one electron in a field of others) using approximations in both constructing the wavefunctions as well as the Hamiltonian, the energy operator. The guess wavefunction is constructed as a linear combination of atomic basis functions and optimized using empirical approximate methods such as Huckel [13] or Zero Differential Overlap (ZDO) [14] between atomic functions. This results in an conditional eigenvalue problem, which requires orthonormalization of the wavefunction, and the diagonalization of the Fock (energy operator) matrix, which consists of integro-differential terms from the Hamiltonian due to the inclusion of one and two electron interaction terms and the differentials with respect to spatial coordinates. The diagonalization results in the eigenvalues (energies) and eigenvectors (the molecular orbitals) that describe the molecular system. The scaling of these algorithms varies as $N \log N$ to $N^7$ depending on how many electronic configurations (required for excited states) are considered, where $N$ is the number of basis functions. The improvements of the wavefunction results in lower energy and the iterations continue until the energy can not be reduced further. This procedure occurs at a given molecular geometry (3D configuration of atoms) and can be repeated at modified geometric configurations and the energy (and the wavefunction) can be further optimized with respect to the geometry. Broadly, the approximations include those in the construction of initial wavefunction in terms of number and types of Gaussian functions to be used to describe the systems and in constructing the Hamiltonian in terms of number and type of operators used.

*2.2.1 Gaussian Input.* A Gaussian input file consists of a Route section containing keywords that describes the computation, Title section, and the Molecule specification that defines the molecular configuration. In addition, a system specification section (aka link 0 section) that may describe computer parameters such as memory, processors and a checkpoint file for data reuse with route options for restart or data reuse. The Route section specifies the desired calculation or job type such as energy at the given configuration (single point or SP), geometry optimization (Opt), reaction path (IRC), harmpnic normal mode analysis and thermochemical analysis (Frequency), and the model chemistry (for exampe G1-G4 etc). Different job types can be linked using link keywords or multiple route keywords to create a chain of executions, for example, calculating frequency after a geometry optimization run or frequency analysis after SP calculation. These chained and multistep runs specified in the input file lead to successive computations. We plan to address the advanced execution cases in future work. The following are some of the route keywords we extracted from the Gaussian input files for our calculations:

- *Basis sets.* These are mathematical functions that describe the atomic electronic wavefunctions that are combined to form molecular orbitals. There are many approximations of these Gaussian functions that represent an exponential function called *Slater type orbital* that come in many variations: Double-Zeta, Triple-Zeta, Quadruple-Zeta, Polarized Sets, Split-Valence, and Diffuse Sets etc., that differ in the number of exponents and corresponding coefficients of a Gaussian function of the form $e^{-c\sigma^2}$ for the combinations needed to provide the approximations.
- *Methods.* Quantum chemical calculations are conducted under various theoretical models such as Molecular Orbital Theory, Density Functional Theory, Perturbation Theory, Coupled Cluster theory, and provide different levels of accommodating electron correlation and inclusion of excited electronic states.
- *Solvents.* Molecular systems can be studied in gas phase or in a solvent again using various models to represent the solvent.
- *Time-dependent.* The fundamental solution for these systems comes from solving the Schrodinger's or Kohn-Sham's equation which has two forms – a time independent and a time dependent form. The default form is time independent. However, an input methods keyword to specify time dependent model can be set with variations and additional options.
- *Intrinsic Reaction Coordinate.* It is a calculation that enables the computation of a chemical reaction that connects the transition state geometries of the reaction and ground state geometries of the reactant and product using a normal mode following intrinsic reaction coordinate (IRC) calculation.
- *Optimize.* The program can optimize the geometry of a molecule (3D configuration) for which the molecular configuration can be defined in three different types of coordinate systems: (1) a z-matrix, which provides a internal coordinate system starting with a reference atom (2) redundant or mixed system, and (3) Cartesian system with various constraints.

## 3 SIMULATION DATASET

### 3.1 Data Origin

*3.1.1 Input Data Collection:* SEAGrid [7] is a science gateway built to execute experiments such as computational chemistry, molecular dynamics, and fluid dynamics. Some of the XSEDE HPC systems used by SEAGrid include the following: Bridges, Comet, Wrangler, Stampeed2 (current), and Blacklight, Trestles, Gordon, and Stampede (past). Gaussian jobs run through the SEAGrid gateway are logged with the input along with the job metadata files. For this analysis we used the Gaussian dataset collected based on runs performed on SEAGrid Gateway from 2005 to 2015 under GridChem middleware [21],[18]. We will consider the data collection that changed under newly deployed Apache Airavata middleware after 2015 in the future.

In Table 1, we present the route keywords, system, and derived features as the independent variables used in our learning models. As discussed previously in the Background section, route keywords are Gaussian input file instructions that specify the job type on Gaussian. In addition to the route keywords, we also obtain system features of the cluster running Gaussian application. Also, we include derived features computed from the molecular specification section of the Gaussian input file.

The system and derived features are the following:

- *Normal Termination.* It is the status message in the output (log) of an an experiment and is used to validate the successful completion of a Gaussian run. If it is not found in the log, a restart run with reuse of data from the first could have been executed to complete the run and a combined run time from both runs is considered.

- *System.* It represents the hardware system, typically identified by a name, in which the execution of the experiment was carried out. The hardware systems change as the resource providers retire and introduce new systems for service.
- *Charge.* The software can treat neutral or charged molecular systems, or radical ions and species, with different total electronic spin.
- *Atoms.* Molecules consist of atoms and the number and type of atoms critically influence the run time.
- *Electrons.* The number of electrons is dependent on the atoms and their atomic number and the overall charge.
- *Run Time.* The run time is the independent variable, which is the property to be estimated and is influenced by many parameters described above as well as the hardware and network properties and the environment in which the Gaussian software is executed.

### 3.2 Pre-processing

*3.2.1 Missing values.* Due to the wide-ranging capabilities of Gaussian, our collected data has different route keyword specifications requiring extensive pre-processing. In order to use the data in our different machine learning models, our data pre-processing techniques include filling missing data with default values for categorical, and binary or boolean variables. We set their respective default values as described in the Gaussian user manual [6].

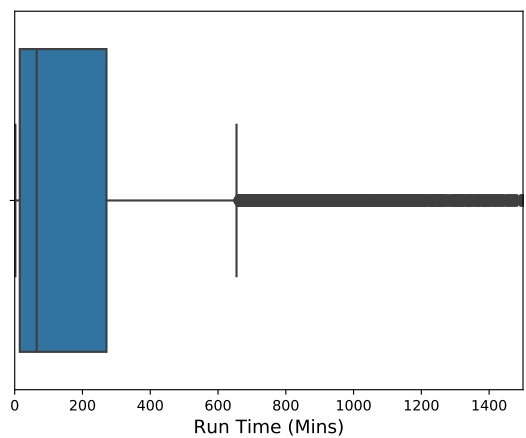

**Figure 1: Box plot of run times.**

*3.2.2 Remove Erroneous Data.* In Figure 1, we show a box plot of the run times. It highlights three cases that are highly unusual with run times between 1 month and up to 571 days. These values could be from compounded runs with multiple restarts or suffering convergence errors due to the user misconfiguration or inherent in the system itself. These data points are omitted and will be reconsidered in future.

*3.2.3 Data Coding.* Various route and system keywords are categorical in nature, which poses a significant issue when we wish to compare runs with different categorical inputs. In order to be consistent throughout our regression models, we encoded our categorical route keywords in one-hot vector [36]. We used the one-hot encoding technique to create a binary column for all our categorical values or a binary representation of their different options. In Table 1, we list all the categorical and Boolean route keywords (i.e., basis sets, methods, and solvents) as well as the system keywords for which we used one-hot encoding.

| Keywords/Features | Name | Type |
|---|---|---|
| System Features | normal termination | Boolean |
| | system | Categorical |
| | run time | Float |
| | numprocs | Numeric |
| Route Keywords | basis set | Categorical |
| | method | Categorical |
| | restricted unrestricted | Boolean |
| | solvent | Categorical |
| | time dependent | Integer |
| | Intrinsic reaction coordinate | Boolean |
| | optimize | Boolean |
| Derived Features | charge | Numeric |
| | multiplicity | Numeric |
| | atoms | Numeric |
| | Electrons | Numeric |

**Table 1: Keywords and Features:** *A subset of keywords use in the dataset. Most of the keywords are extracted from the input file provided by the user for the Gaussian execution. Route Keywords are the actual inputs that Gaussian needs for execution. Derived features are not explicitly in the Gaussian input or SEAGrid database, but were computed from those to provide minimal information about the initial molecular structure.*

## 4 EXPLORATORY DATA ANALYSIS (EDA)

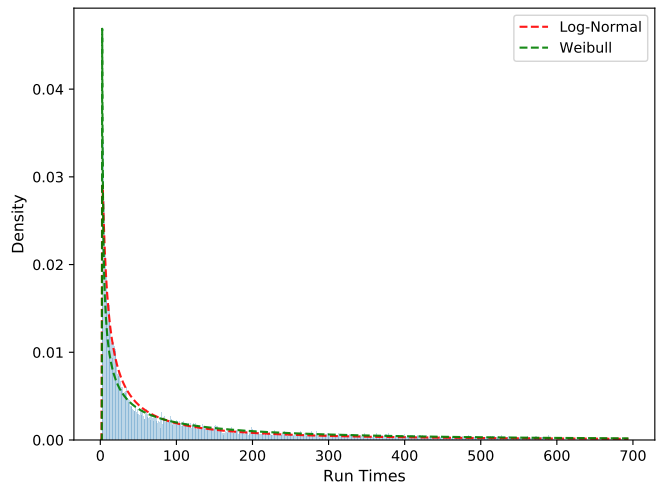

**Figure 2: Run Time:** *Histogram of the density of run times for 700 minutes with the Log-Normal and Weibull distributions after MLE. The univariate analysis shows the majority of the run times are below 700 minutes. Hence, we truncate the histogram and distributions to better illustrate the trends.*

We initially performed some exploratory data analysis and statistics to extract information on the characteristics of the dataset. In Table 2, we show the standard central tendency measures, and generated the run time histogram in Figure 2. We notice that even though most of the run times between 2 to 200 minutes, the mean is 270.2 minutes. The standard deviation of 606.54 minutes also suggests that the run time is heavily tailed distribution in our current dataset. We also see that the change in range increases by only 134 minutes from 25 % to 75% of the data, furthering corroborating that, a small number of high time consuming runs make up the majority of the run time in the dataset.

We performed maximum likelihood estimation (MLE) on the run times with Weibull [4]and Log-Norm [29] distributions to further analyze the run time trends in our dataset. Following the MLE, we obtain the log-normal distribution,

$$f(x) = \frac{k}{s(x-x_0)\sqrt{2\pi}} exp\left(\frac{-log^2\left(\frac{x-x_0}{k}\right)}{2s^2}\right) \quad (1)$$

with s = 2.06, $x_0$ = 1.95, and k = 55.19. We also obtain the Weibull distribution,

$$f(x) = s\left(\frac{x-x_0}{k}\right)^{s-1} exp\left(-\left(\frac{x-x_0}{k}\right)^s\right) \quad (2)$$

with s = 0.54, $x_0$ = 2, and k = 152.49. We also performed the Kolmogorov-Smirnov (KS) [16] Test to judge the goodness of fit of these distributions. The KS Statistic for the run times with the above Log-Norm distribution is 0.033 with a p value of $5.67 \times 10^{-36}$. The KS statistic for the run times and the Weibull distribution given is 0.051 with a p-value of $9.72 \times 10^{-84}$. Both distributions fit well to the run time data. Both the MLE and KS Test were performed using the SciPy library [15]. As we use the route keywords from the Gaussian input files as independent variables, we run some statistics on the basis sets and methods used in our dataset to understand their distribution. The basis sets and methods are integral to the description of the job type in Gaussian. Hence, we generated the histograms for the basis sets and methods in our data. In Figure 3a, we can observe that the top 5 most frequent basis sets contain more than 70% of the data. Similarly, in Figure 3b, we show that the default method *b3lyp* makes up 43% of the dataset.

## 5 REGRESSION METHODS

The Gaussian route keywords determine the type and number of calculations being performed by the software. Therefore, we can map our run time prediction problem as a canonical regression problem, using the route keywords, derived keywords and system keywords as the input features. Since, we assume no prior knowledge as to the correlation of the input features, we tried a number of different regression models, results from which are presented in the Table 3. We used 10-fold cross-validation. All the experiments in Section 5.1 and 5.2 were performed using using scikit-learn [19] and TensorFlow [2]. Default values were used for any configuration of hyper-parameters.

### 5.1 Linear Methods

We initially performed linear regression on the entire dataset and used the results as the baseline. As can be seen in Table 3, the error was generally high. The initial linear models achieved over 530 minutes of Root Mean Square Error (RMSE). To identify other linear models that can fit the data, we also tried Ridge Regression (RR) [33] and LASSO regression (LLR) [32] both yielded a RMSE of 532.19 minutes and 541.68 minutes respectively. RR calculates error with the sum-square of residuals where as, LLR calculates it with the sum of absolute values of the residuals. Both RR and LLR perform within similar bounds as to those of LR. Given that we obtain similar results for the previous linear regression models, we execute Elastic-Net Regression (ENR) [31] model, which combines both minimization techniques from RR and LLR. In Table 3, ENR performance degraded from 28 to 40 minutes above the other linear regression models. Since, there was not any improvements in the linear models, the linear combination of route keywords, system and derived features as presented is not able to predict the run time. We observe that there is no significant improvements in any of the linear models, and an average performance degradation of about 21 minutes for LLR and ENR.

### 5.2 Nonlinear Methods

Given the high RMSE of the linear methods, we next tried a number of nonlinear methods. The nonlinear models include K-Nearest Neighbors ($k$-NN) [35], Regression Tree (RT) [37], and a Deep Neural Network (DNN) [10]. DNNs are able to model complex non-linear relationships and the layered structure is able to compose features from a lower hidden layer. Our DNN was created using the Tensorflow Estimator API [1], had 3 hidden layers, with 1024, 64, and 32 nodes respectively. We used the Sigmoid activation function per layer to introduce non-linearity and batch normalization per layer to improve training efficiency. We used a 32-element mini batch technique during training alongside the Adam Optimizer and a learning rate of $\epsilon = 10^{-3}$. We also used dropout regularization of 0.7 per layer as well.

The $k$-NN method identifies $k$ number of instances in the dataset for a particular new data instance. The respective mean is taken and used as a prediction from the $k$ neighbors identified. In the particular $k$-NN implementation we use the default metric where k=5 and the distance measurement of neighbors used is Minkowski distance [34], which is a generalization of both Euclidean distance and Manhattan distance. In Table 3, we show that among the nonlinear subset of models, DNN outperforms *k*-NN by 12 minutes. However, there was a performance improvement of over 40 minutes for both *k*-NN and DNN when compared to linear regression models. The regression tree model selects the best points to split the data to minimize the cost function. The default cost metric for RT is the mean squared error. Moreover, the results shown in Table 3 point to the RT model under-fitting the data.

| Features | Data Points | Mean | STD | Minimum | 25% | 50% | 75% | Maximum |
|---|---|---|---|---|---|---|---|---|
| *Electrons* | *37,034* | *195.45* | *184.64* | *0* | *104* | *168* | *238* | *10,877* |
| *Atoms* | *37,034* | *37.13* | *24.06* | *0* | *19* | *36* | *49* | *315* |
| *Run Time (Minutes)* | *37,034* | *277.21* | *606.54* | *2.00* | *14.90* | *64.90* | *270.79* | *24,804.28* |

**Table 2: Statistics of current dataset for 2 independent variables (i.e., atoms and electrons) and the dependent variable (i.e., run time). This graphs only shows a subset of parameters as the other parameters did not provide any statistical significance.**

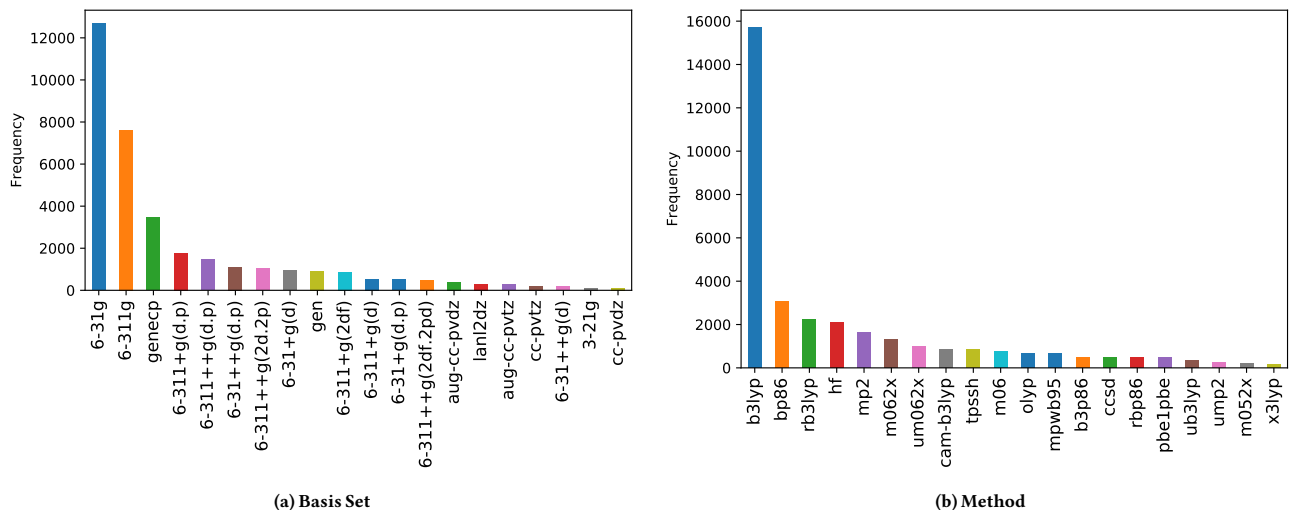

(a) Basis Set          (b) Method

**Figure 3:** *Histograms of the top 20 most frequent items 3a basis sets and 3b methods. 3a presents the frequency of basis sets from 106 for 6-31+g(2d.p) to 13163 for 6-31g. In 3b, we have the smallest methods ranging from 190 (x3lyp) to 16200 (b3lyp) instances.*

| Type | Model | RMSE(Min) |
|---|---|---|
| Linear | Linear Regression (LR) | 533.06 |
| | Ridge Regression (RR) | 532.19 |
| | LASSO Linear Regression (LLR) | 541.68 |
| | Elastic Net Regression (ENR) | 567.35 |
| Non-Linear | K-Nearest Neighbors (KNN) | 494.09 |
| | Regression Tree or CART (RT) | 540.16 |
| | Deep Neural Network (DNN) | 482.85 |
| Ensemble | Random Forest (RF) | 478.24 |

**Table 3: Linear, Non-Linear, and Ensemble RMSE.**

The different hyper-parameters in the non-linear model did not have a large impact when fitting the data. We tuned the topology of DNN to identify if deep, wide, or a combination of both topologies will yield better results. Our results show that changing the topology of DNN does not yield better results.

Finally, we performed regression evaluations with Random Forest (RF) [30]. RF is a set of decision trees. Each decision tree considers a random subset of route keywords and features. In Table 3, RF outperforms all the chosen models. However, since its performance is fairly close to the performance of the DNN, we make use DNN for all the experiments in section 5.3. Unlike linear and non-linear models, RF was able to make a slightly better generalization over the entire dataset.

### 5.3 Data Subsetting

Regression over the entire data set performed unsatisfactorily. In order to better understand why this was the case, we decided to focus on specific subsets of the data. Some of the subsets are based on the dependent variable (run time), and thus could not be used in actual prediction; since the run time is not available beforehand. Thus, we present those results solely as an exploratory result. Other subsets are based on the input features, and thus could be used in practice.

*5.3.1 Run Time Percentiles.* In this analysis, we attempt to ascertain whether or not it was a particular subset of the jobs, based on run times, that was particularly hard to predict, resulting in high RMSE. So, we ran regression on runs in the 25th (below 14.9 minutes), 50th (below 64.9 minutes), and 75th (below 270.75 minutes) percentiles as shown in Table 4. When including only runs in the 25th percentile, we achieved RMSEs of 4 minutes for LR and DNN respectively. For runs in the 50th percentile, we achieved RMSEs of 17 minutes and 16 minutes for LR and DNN respectively. For runs in the 75th percentile, we achieved RMSEs of 63 minutes and 90 minutes for LR and DNN respectively.

*5.3.2 Mode of Atoms.* In order classify the importance of the difference class of inputs, we created a new subset by holding the number of atoms constant in the data set. We computed the mode of the number of atoms of the runs in the data set, and ran regression on those runs separately. We observed that there was no significant improvement. In Table 4, we show that the standard deviation, for the subset based on the mode of the number of atoms, is 9% higher than the whole dataset (see Table 2). The subset of mode of the atoms represented only 3% of the dataset, which shows that the values have a high variation over a small subset of the data. Moreover, the mode presented an RMSE increase of 10% and 20% in comparison

PEARC '19, July 28-August 1, 2019, Chicago, IL, USA                                                                                                                                          Beltre et al.| Dataset | Datapoints | Run Time (Minutes) | | | RMSE (Minutes) | |
| --- | --- | --- | --- | --- | --- | --- |
| | | *Median* | *Mean* | *STD* | LR | DNN |
| **25th Percentile** | 9259 | 6.45 | 7.2 | 3.64 | 4 | 4 |
| **50th Percentile** | 18518 | 14.9 | 20.26 | 16.99 | 17 | 16 |
| **75th Percentile** | 27777 | 30.12 | 60.43 | 66.99 | 63 | 90 |
| **Atoms Mode** | 1290 | 16.21 | 229.15 | 662 | 591 | 583 |
| **Electrons Mode** | 1071 | 13.57 | 39.62 | 70.26 | 73 | 77 |
| **Atoms and Electrons Mode** | 354 | 7.87 | 12.02 | 12.72 | 13 | 15 |
| **Constant Route Keywords** | 548 | 81.16 | 130.55 | 146.06 | 106 | 109 |
**Table 4: Data Sub-setting:** *A summary of data subsets based on percentiles, mode of derived features, and Route keywords.*

| Dataset | Datapoints | Run Time (Minutes) | | | RMSE (Minutes) | |
| --- | --- | --- | --- | --- | --- | --- |
| | | *Median* | *Mean* | *STD* | LR | DNN |
| **25th Percentile** | 9259 | 6.45 | 7.2 | 3.64 | 4 | 4 |
| **50th Percentile** | 18518 | 14.9 | 20.26 | 16.99 | 17 | 16 |
| **75th Percentile** | 27777 | 30.12 | 60.43 | 66.99 | 63 | 90 |
| **Atoms Mode** | 1290 | 16.21 | 229.15 | 662 | 591 | 583 |
| **Electrons Mode** | 1071 | 13.57 | 39.62 | 70.26 | 73 | 77 |
| **Atoms and Electrons Mode** | 354 | 7.87 | 12.02 | 12.72 | 13 | 15 |
| **Constant Route Keywords** | 548 | 81.16 | 130.55 | 146.06 | 106 | 109 |

**Table 4: Data Sub-setting:** *A summary of data subsets based on percentiles, mode of derived features, and Route keywords.*

to the entire dataset for LR and DNN respectively. We were unable to reduce the prediction error compared to the standard derivation. Since the molecular specification may provide more quantitative information for certain categorical features, further specialization of our input features may be required to improve the prediction.

*5.3.3 Mode of Electrons.* We also tried holding the number of electrons constant. We used the subset where the number of electrons equalled the mode of the number of electrons over the entire data set, in order to maximize the number of data points in our subset. We obtained 1071 records, which is 17% fewer records than the mode of Atoms. In Table 5.3, we present that the standard deviation for a constant number of Electrons is 70.26 minutes. Also, we present that the RMSE is 73 minutes and 77 minutes for LR and DNN respectively. For both linear and non-linear models, we were able to improve over RMSE. The *mean* for Electrons subletting is 39.62, which is substantially lower than the mean yielded by subletting Atoms. As a result, we observe that holding Electrons constant based on the mode improves the overall accuracy of the model slightly, indicating possible predictive power for the derived keywords in our input. This is also in line with theorist's intuition that electrons are more fundamental and influence the runtime more directly.

*5.3.4 Mode of Atoms and Electrons.* Finally, we held both the number of atoms and the number of electrons constant. Using the mode of the combination of both, in Table 4, we show that a small subset of 354 records were captured when keeping both derived features constant. The overall RMSE time was reduced to 13 minutes and 15 minutes for LR and DNN respectively. These RMSEs are lower than previous results presented, but given that it is not below the standard deviation of about 12.72 minutes, it shows that both models cannot demonstrate the ability to learn.

*5.3.5 Constant Route Keywords.* The results shown by the preceding experiments were not sufficient to enable the models to learn. Next, we performed regression by keeping the route keywords constant and allowed derived keywords to change. By keeping the route keywords constant, we extracted a data subset of 548 records. In Table 4, we show that the RMSE is 106 minutes and 109 minutes for LR and DNN respectively. The standard deviation for the data subset demonstrates the ability to learn. However, as we carried out further explorations to overfit the model with the training set, both models were unable to improve learning.

## 5.4 Feature Specification

The preceding results suggest that the input features taken directly lack predictive power. In other words, there is simply not enough discriminating information for the machine/algorithm to learn from our set of independent variables. Further investigation showed that, indeed, using the route keywords, job meta data, and derived features by themselves do not provide sufficient explanatory information. For example, the mode of all the input features combined appeared 300 times in the dataset and the run times of varied from 4.83 to 125.4 minutes with a variance of 619.38. Thus, regardless of the run time predicted for that run using our current features, the error will be high.

This implies that additional quantitative description of the actual molecular and other feature specifications are important. This is not surprising, but our exploratory results confirm this. Future work will examine the molecular specification in terms of adjacency matrices and quantitative description of features such as basis functions in terms of primitives and method in terms of algorithm complexities as independent variables.

## 6 RELATED WORK

Predicting the execution time of large scale programs has been an longstanding area of computer science research. Our investigation aims to identify useful subset of input parameters to estimate run time of such applications. Various statistical and machine learning methods have been proposed to model execution time and understand the features contributing to run times. Huang et al. [12] introduced a Sparse POlynomial REgression (SPORE) methodology, to build accurate prediction models and select the most relevant subset among a few hundred input features using a sparse and non-linear model by determining input features that dominate the execution time. The model was tested with image data with highly varying run time, but seemingly homogeneous inputs and was able to achieve state of the art accuracy.

The underlying architecture of the system can have a significant effect on execution time of programs. Finkler and Mehlhorn [8] analyzed the problem of predicting the run time of the same program on varying micro-architectures. The two main methods discussed for analysis are regression analysis and operation counting. Regression analysis of programs required obtaining the constant values of the asymptotic behavior of the algorithms involved. Even for simple cases, the architecture, caches, and pipelines caused a large variance in run time. The systematic errors inherent in the model are difficult to overcome. Operation counting is a technique that counts the number operations and by proxy the number of cycles a program requires. This method requires low-level access to the programs and involved overloading various CPU operations in order to track and predict run time.

In recent years machine learning approaches to run time calculations for large programs have been proposed. Gupta et al. [11] [5] used a variant of decision trees named Predicting Query Runtime (PQR) tree to predict the execution time of database queries. The author predicted the execution time of database queries with various load conditions and differing input features of the queries that have a simplistic relation with execution time. For coarse-grained approximations of run time, decision trees have been shown to perform well, if the relation between the inputs to the PQR and the run-time is sufficiently simple. Matsunaga and Fortes [17] build on top of the PQR tree proposed by Gupta et al. to predict the run time of bioinformatics applications BLAST and RAxML using machine learning techniques. They also explored the use of two other learning algorithms, support vector machines (SVM) and k-nearest neighbor, to predict execution time. They proposed the Predicting Query Run time Regression (PQR2) algorithm, a generalization of the classification of tree approach of PQR, to produce fine-grained predictions. PQR2 consists of a tree with each leaf containing a regression that is chosen from a pool. The linear and SVM regressions provided good results as a result of the linear relationship between the length of the input sequence and execution time.

Modelling execution time of disparate programs in varying systems and architectures is a difficult task due to the multiplicity of factors that may affect running time. Yet, as many large scale programs share similar underlying algorithms, they lend themselves to a similar course of study. Popescu et al. [20] proposed a methodology to predict the run time of a class of large scale iterative algorithms. Predicting the execution time for Gaussian calculations have thus far depended on the algorithm and CPU performance counters and not the input parameters. Anthony et al. [3] proposed a linear performance model (LPM) to model the run time of the Gaussian Computational Chemistry Software for Hartree-Fock (HF) and hybrid Hartree-Fock for Density Functional Theory methods. The number of computation cycles is calculated using the instruction count and L1 and L2 cache misses for varying architecture. The cycle count of HF and hybrid HF/DFT calculations primarily depend on the computation of two-electron repulsion integral (ERI) and the PRISM algorithm used to compute the integral. Two basis sets are used for bench-marking. Gaussian contains more sophisticated computational methods based on HF with varying computational complexity. The model relies on hardware performance counters and costly cache simulation mechanism to model run time. As computation based research is driving scientific discoveries, robust, efficient and easy to use infrastructure is getting significant attention from the community. Estimating run time of scientific experiments is important to utilize the cluster resources better in a multi-tenant setup. Cloud computing research [25] [26] [27] [28] has shown how modern cloud technologies can be useful in setting up multi-tenant cloud environment to host scientific computing workloads [22].

## 7 CHALLENGES

Based on our experience of conducting this research study, we present the challenges in predicting the run time of Gaussian Chemistry code.

(1) *Challenges with the dataset.* The current state of dataset extraction process can be tuned to enable a seamless data extraction ecosystem. One of the main challenges faced by the current data extraction process is the number of heterogeneous systems in which the data is generated. In order to enhance the platform and create new results, storing the input file directly in the database will enable the users to extract more data and re-engineer features from the large sets.

(2) *Challenges in distributed systems and shared clusters.* The data represented here spans a decade or longer and during this time many HPC systems that were in production have been retired. So, the results need to be mapped to the current systems in operation to be useful. The same application is also deployed in different systems and the recommendation is different for each. Users may want a total time to solution, wherein run time is only one component. We have other solutions, such as queue wait time prediction for a HPC systems that need to be combined and presented for all possible systems to the users. Software upgrades and new revisions can also change the performance, run times, and also introduce novel functionality. Prediction networks trained previously may not provide accurate estimations of run times for the current state of the software and target infrastructure.

## 8 CONCLUSION AND OUTLOOK

This paper offers insights based on statistical analyses to aid the understanding of Gaussian input parameters. We use domain specific knowledge to characterize the Gaussian's specific tasks. We were able to reduce the feature space by using traditional statistical methods. We show that the qualitative feature representation is insufficient to provide reliable learning by the models we used. Different regression models were not effective in developing insight on the data. In addition, the results show that models that are used to extract higher order embedded features from the dataset performed worse than the models that do not use any hyper-parameters, except RF. As our findings led us to the exploration of basis sets, the number basis-functions may need to be utilized for a better estimation of run time along with the detailed description of the molecular structure. The different RMSE computed for the various models highlights the need for better extraction of derived features from input parameters, and the optimization of the different hyper-parameters.

This work was primarily focused on the exploration of the Gaussian input file parameters directly. Our future work will investigate the runtime prediction using quantitative molecular structure in terms of adjacency matrices, basis sets as primitive functions and other derived quantitative data based on the input route keywords as well as system parameters such as processed instructions per unit time, memory bandwidth and I/O performance for more comprehensive learning.